\documentclass[12pt, letterpaper]{article} 
\usepackage[onehalfspacing]{setspace}
\usepackage [english]{babel}  
				
\usepackage[numbers]{natbib}

\usepackage[margin=1in]{geometry}
\usepackage[mathscr]{euscript}
\usepackage{amssymb}
\usepackage{amsthm}
\usepackage{amsmath}
\usepackage{mathtools,eqparbox}
\usepackage{amsfonts}
\usepackage{csquotes}
\usepackage{epstopdf}
\usepackage{graphicx}
\usepackage{subcaption}
\usepackage{stackengine}
\usepackage{multirow,array}
\usepackage{booktabs}
\usepackage{bm}
\usepackage{comment}
\usepackage{float}
\usepackage{booktabs}
\newcommand*\vect[1]{\mathbf{\bm{#1}}}

\usepackage{xpatch}
\usepackage{hyperref}
\hypersetup{colorlinks=true,citecolor=blue}

\title{Bayesian Inference for Confounding Variables and Limited Information}
\author{Ellis Scharfenaker\thanks{Department of Economics, University of Utah, 260 Central Campus Dr, Gardner Commons, Salt Lake City, UT 84112, US. (ellis.scharfenaker@economics.utah.edu)}\and Duncan K. Foley\thanks{Department of Economics, The New School for Social Research, New York, NY, USA}}

\begin{document}

\date{\today } \maketitle

\begin{abstract}
\singlespacing
A central challenge in statistical inference is the presence of confounding variables that may distort observed associations between treatment and outcome. Conventional ``causal" methods, grounded in assumptions such as ignorability, exclude the possibility of unobserved confounders, leading to posterior inferences that overstate certainty. We develop a Bayesian framework that relaxes these assumptions by introducing entropy‑favoring priors over hypothesis spaces that explicitly allow for latent confounding variables and partial information. Using the case of Simpson’s paradox, we demonstrate how this approach produces logically consistent posterior distributions that widen credibly intervals in the presence of potential confounding. Our method provides a generalizable, information‑theoretic foundation for more robust predictive inference in observational sciences.
\end{abstract}


\section{Introduction}

All problems in statistics share a common difficulty: incomplete information.  Data may omit relevant variables, capture only aggregated summaries, or reflect structural dependencies that are only partially known.  Standard inferential methods often respond to this incompleteness by imposing strong assumptions about unobserved confounding that simplify analysis but restrict the hypothesis space. Such assumptions are rarely testable in practice, and as we show below lead to posterior distributions that understate uncertainty.  

We propose a general Bayesian framework for inference under incomplete information. The key idea is to use \emph{constrained entropy-favoring} (CEF) priors over hypothesis spaces, which explicitly allow for unobserved or latent factors while incorporating whatever partial information is available through maximum-entropy constraints.  This yields posterior distributions that are more cautious when information is scarce, but systematically sharpen as constraints are added. The approach is general, encompassing sensitivity analysis with unmeasured confounders, missing-data problems with partially observed distributions, and more broadly any setting where the structure of dependencies is only partly known.  

Confounding variables provide a natural and important application. They remain a central challenge for statistical inference, as they can amplify, reverse, or obscure associations between treatment and outcome. Many widely used ``causal'' methods proceed under strong, untestable assumptions---most notably the absence of unmeasured confounders (the ignorability assumption; \citealp{rubin2015}). While simplifying analysis, this assumption restricts attention to models without hidden confounding and can yield posterior distributions that overstate certainty. Unobserved confounding remains a major obstacle in statistical practice \citep[ch. 9]{hardt_recht2022}. 

Incomplete information about confounders is also common in observational sciences. 
Clinical trial data, for example, may record only treatment and recovery, omitting relevant characteristics such as sex, age, or socioeconomic status, or provide only partial summaries such as marginal or conditional distributions. Analyses based on different levels of information can lead to different effect estimates.  

Our CEF framework relaxes these assumptions by explicitly incorporating possible unobserved confounders and partial information, ensuring that posterior distributions widen appropriately when information is limited and adjust systematically when only aggregate or incomplete data are available. CEF priors translate theoretical or summary knowledge into usable constraints on the hypothesis space, preventing the exclusion of valuable but imperfect information. The framework also supports sensitivity analysis, allowing researchers to assess how confounders of varying strength could alter posterior inferences.  

From the Bayesian perspective, the informational structure of hypotheses need not be restricted by the informational structure of the data. Information theory \citep{Cover2006, Golan2018, Jaynes_1957b} provides a rigorous foundation for embedding such prior information into statistical models by maximizing Shannon entropy subject to specific informational constraints \citep{Jaynes1968, Jaynes1983}. The Constrained Maximum Entropy (CME) framework has demonstrated substantial utility across a range of applications, including machine learning \citep{Gupta2006, Murphy2024}, classification \citep{Fujino2008, Chau2001}, interaction networks \citep{Lezon2006}, clustering \citep{Beni1994, Qjidaa1999}, and advanced image processing tasks such as enhancement and denoising \citep{Skilling1984, Smith1979, Henter2016, Rioux2021}, disease detection \citep{Zadran2014, Remacle2010}, and predictions in biological systems for medical interventions \citep{Young2011, Kravchenko2020, Flashner2019}.  

We demonstrate the benefits of this approach in predictive settings through the canonical example of Simpson's paradox, in which treatment appears harmful within subgroups but beneficial overall \citep{Lindley1981}. Our framework integrates partial information transparently, producing posterior inferences that remain logically consistent and appropriately cautious when confounders are observed, unobserved, or partially observed.  

\section*{Simpson's paradox and its implications}

Consider a clinical trial of $N=80$ participants, half treated and half not. The outcome is whether each participant recovered. Table~\ref{table:SPLimited} shows that the treated group had a $50\%$ recovery rate compared to $40\%$ for the untreated, suggesting a beneficial effect of treatment under the assumption of exchangeability \cite[Ch.11]{definetti1974}.

Now suppose we also know participants’ sex. Treating the data as \textit{partially exchangeable} within sex reveals a different picture (Table~\ref{table:SPFull}): untreated males and females both recover at higher rates than those treated. This apparent reversal is Simpson’s paradox \cite{novick1983, pearl2014}.

The paradox raises two questions: what if this additional information were unavailable, and how can we be sure there is not some other unobserved confounder? While we can never be certain, Bayesian inference allows us to incorporate the possibility of such confounders into the hypothesis space. Doing so increases the caution of posterior inferences without necessarily changing point estimates.

\begin{table}[ht!]
\centering
\caption{Hypothetical contingency table for treatment and outcome. With no further information, these data may be treated as exchangeable.}
\label{table:SPLimited}
\begin{tabular}{lcccc}
\toprule
 & Recovered & Not Recovered & Total & Recovery Rate \\
\midrule
Treated      & 20 & 20 & 40 & 0.50 \\
Not Treated  & 16 & 24 & 40 & 0.40 \\
\bottomrule
\end{tabular}
\end{table}

\begin{table}[ht!]
\centering
\caption{Hypothetical contingency tables by sex. With sex information, the data are partially exchangeable.}
\label{table:SPFull}
\begin{tabular}{lcccc}
\toprule
 & Recovered & Not Recovered & Total & Recovery Rate \\
\midrule
\multicolumn{5}{c}{\textbf{Males}} \\
Treated      & 18 & 12 & 30 & 0.60 \\
Not Treated  & 7  & 3  & 10 & 0.70 \\
\midrule
\multicolumn{5}{c}{\textbf{Females}} \\
Treated      & 2  & 8  & 10 & 0.20 \\
Not Treated  & 9  & 21 & 30 & 0.30 \\
\bottomrule
\end{tabular}
\end{table}
In this paper we argue for the following points:

\begin{enumerate}
\item Treatment and other consequential decisions are best based on Bayesian posterior probabilities derived from constrained entropy-favoring priors over well-defined hypotheses and all available information, including characteristics of trial subjects.
\item Different decision-makers with different information or priors may reach different conclusions. As De Finetti \cite{definetti1974} emphasizes, probability theory ensures the \textit{consistency} of inferences, not their conformity to some ideal truth.
\item While the information contained in the data constrains the likelihood, it does not constrain the hypotheses considered. Even without direct data on a confounder, one can include its possibility in the prior.
\item In the absence of actual data on any particular confounder, including the possibility of confounders in the hypothesis space does not change modal posterior estimates but widens distributions, reflecting a lower confidence in the conclusions. Constrained entropy-favoring priors embed caution absent from approaches that ignore unobserved confounders.
\item The Bayesian approach also allows incorporation of partial or theoretical information directly into posterior inference through constrained entropy-favoring priors.
\end{enumerate}

By grounding predictive inference in information theory, our method provides a generalizable foundation for analyzing complex observational data across the sciences, offering a principled alternative to conventional ``causal'' frameworks that rely on restrictive assumptions.

\section{State spaces, data, and the likelihood}

Consider a system of $M$ identical components, each observed in one of $K$ possible outcomes. We denote the outcome set by $\mathcal{X}$. A microstate of the system is the outcome vector $s=\{x_1,\ldots,x_M\}$ with $x_m\in\mathcal{X}$, and the microstate space is $\mathcal{S}=\mathcal{X}^M$.

If the system is \textit{exchangeable} in De Finetti's sense \citep{definetti1974}, its macrostate can be described by the normalized histogram of counts across outcomes, $\vect{q}=\Big\{q_1=\tfrac{m_1}{M},\ldots,q_K=\tfrac{m_K}{M}\Big\}$ where $m_k$ is the number of components in outcome $k$ and $\sum_k m_k=M$. A Bayesian hypothesis is then a frequency assignment $\vect{q}$ with $\sum_{k=1}^K q_k=1$. When the hypothesis space is higher dimensional, we may equivalently represent it in terms of conditional or joint frequencies.

Given $N$ exchangeable observations, the data are summarized by the counts $\vect{n}=\{n_1,\ldots,n_K\}$, or equivalently by the relative frequencies
\(
\vect{p}=\Big\{p_1=\tfrac{n_1}{N},\ldots,p_K=\tfrac{n_K}{N}\Big\}.
\)
We reserve the term \textit{frequency distribution} for macrostates like $\vect{p}$ and $\vect{q}$, which are visualized as points in the $K$-simplex
\(
\mathcal{S}_K=\{\{q_1,\ldots,q_K\}\;|\;q_k\ge0,\ \sum_k q_k=1\}.
\)

The Bayesian perspective distinguishes \textit{probability distributions} that represent an observer's degree of belief about the uncertain state of the system, which we denote $\mathcal{P}$, from frequency distributions that describe a macrostate of the system such as $\vect{q}$. An observer from the Bayesian point of view has a joint prior distribution over the spaces of hypotheses and data, $\mathcal{P}[.,.]:\mathcal{S}_K \times \mathcal{S}_K\rightarrow \mathbb{R}_{\ge 0} \mbox{ with }\int_{\mathcal{S}_K \times \mathcal{S}_K}\mathcal{P}=1$. This joint distribution is often represented as the product of a marginal prior distribution over the hypotheses and a conditional likelihood $\mathcal{P}[\vect{q},\vect{p}]=\mathcal{P}[\vect{p}|\vect{q}]\mathcal{P}[\vect{q}]$.

We adopt the conventions that a function with a vector argument is applied component-wise to the vector and the operator $\cdot$ denotes the dot product of vectors. 

For large $M>>1$ we can write the multinomial likelihood for sample frequencies $\vect{p}$ conditional on the hypothesis $\vect{q}$ as the Kullback-Leibler divergence, or \textit{relative entropy} of the data distribution $\vect{p}$ to the hypothesis distribution $\vect{q}$, which we denote $H[\vect{p}||\vect{q}]\equiv \vect{p}\cdot\log[\frac{\vect{p}}{\vect{q}}]$:

\begin{align}
\label{eq:mult}
\mathcal{P}[\vect{p}|\vect{q}]
&\propto q_1^{n_1} \cdots q_K^{n_K} 
   \propto e^{-N\,\vect{p}\cdot \log\!\left(\tfrac{\vect{p}}{\vect{q}}\right)} 
   = e^{-N H[\vect{p}||\vect{q}]}
\end{align}
With prior $\mathcal{P}[\vect{q}]$, the posterior becomes
\begin{align}
\label{eq:multKL}
\mathcal{P}[\vect{q}|\vect{p}]
&\propto \mathcal{P}[\vect{q}]\,e^{-N H[\vect{p}||\vect{q}]}
\end{align}

Entropy-favoring priors \citep{Jaynes1983,Golan2018,foleygolan2023} weight hypotheses with higher Shannon entropy, which we denote $H[\vect{q}]\equiv -\vect{q}\cdot\log[\vect{q}]$: 
\begin{align}
\mathcal{P}[\vect{q}] \propto e^{H[\vect{q}]}=e^{-\vect{q}\cdot\log[\vect{q}]}
\end{align}
Entropy-favoring priors assign more weight to macrostates with higher multiplicities, adding the least extra information beyond explicit constraints and reducing bias. When theoretical or empirical constraints apply, the prior can be written in constrained maximum entropy form \citep{Golan2018,FoleyScharfenaker2024a}:
\begin{align}
\mathcal{P}[\vect{q}] \propto e^{-H[\vect{q}||\hat{\vect{q}}]}
\end{align}
where $\hat{\vect{q}}$ solves the associated constrained maximum entropy problem.

\section{Predictive inference}

Predictive inference considers what would happen to an outcome, which we represent with $z\in \mathcal{Z}$ as a result of a hypothesized ``treatment'' or intervention $t\in \mathcal{T}$ for some unit, entity, or agent $i$ by comparing the outcomes of the intervention to outcomes in units that do not receive the treatment, often called ``controls''. The fundamental problem of predictive inference concerns the presence of confounding variables, which we denote $a\in \mathcal{A}$. The confounding variable $a$, in effect, represents the summary confounding effect of unmeasured variables. For expositional simplicity we assume that $\mathcal{Z}=\{Z,\bar{Z}\}$, $\mathcal{T}=\{T,\bar{T}\}$, and $\mathcal{A}=\{A,\bar{A}\}$ are all binary variables, as in the Simpson's paradox example. The outcome space is a triple $\mathcal{X}=\{a,t,z\}$, where $z_i$ is the outcome observed for unit $i$, $t_i$ is the treatment indicator for unit $i$, and $a_i$ is an observed or unobserved confounding variable indicator for unit $i$. The hypothesis over the state of the system is a joint frequency distribution $q_{a,t,z}$.

With predictive inference we are often interested in inferring a posterior over the conditional frequency $q_{z|t,a}=\frac{q_{a,t,z}}{q_{t,a}}$, that is the frequency of the outcome conditional on the treatment or intervention and confounding variable. 

If we believe there is no confounding variable $a$,  we can treat the state $\{t,z\}$ as exchangeable, and the joint frequency distribution describing the state of the system is $q_{t,z}=q_{z|t}q_{t}$.

In this case the relationship between the outcome and treatment is expressed in the conditional frequency $q_{z|t}=\frac{q_{t,z}}{q_{t}}$.

We can also express these observations by assigning a numerical value to each one. While it is customary in social and medical sciences to adopt the index $\{0,1\}$ we adopt the indexing more common in physics $\{-1,1\}$, which, as we show below, has more intuitive and convenient mathematical properties. As an example, the treatment variable with this indexing implies the following:

\[
t=\left\{ 
  \begin{array}{l l} 
  	\bar{T} = -1& \text{ if unit $i$ does not receive the treatment } \\
   T = 1& \text{  if unit $i$ receives the treatment } \\
\end{array} \right.\\
\label{eq:index}
\]

The same indexing interpretation applies to outcomes $z$ and confounding variable $a$.
The covariance (and correlation) between treatment and outcome under this indexing defines the weighted difference between the conditional expectations of outcome on treatment, also called the \textit{Average Treatment Effect} (ATE):

\begin{align}
\begin{aligned}
\text{Cov}[t,z]&=\sum_{t\in \mathcal{T}} \sum_{z\in \mathcal{Z}} q_{t,z}t z\\
&=q_{T}\mathcal{E}[z|t=T]-q_{\bar{T}}\mathcal{E}[z|t=\bar{T}]\\
&=\text{ATE}[t,z]
\end{aligned}
\end{align}
Where $\mathcal{E}[z|t]$ is the conditional expectation. 

The ATE for the example in Table~\ref{table:SPLimited} is $0.1$, indicating that the treatment has on average a positive effect on the rate of recovery. The ATE for the data in Table~\ref{table:SPFull} conditional on sex is $-0.2$ for both males and females. While the ATE is a point estimate, we are really interested in a full posterior description of the conditional distribution $\mathcal{P}[q_{z|t}]$, from which we can calculate the \textit{posterior treatment effect} (PTE) as the distribution of the difference of the rate of recovery conditional on treatment $\tau=q_{Z |T} - q_{Z|\bar{T}}$.

Estimating the posterior effect of treatment depends on judgments about the exchangeability of $\{t,z\}$. One approach is to adopt a prior that rules out confounding, effectively restricting the hypothesis space to match limited data, as in Table~\ref{table:SPLimited}. Under this assumption, the positive ATE in Table~\ref{table:SPLimited} represents a consistent Bayesian inference. However, a full posterior distribution provides a more informative assessment, as we show below.

Another important case to consider is when a confounding variable $a$ is part of the joint distribution hypothesis. In this case the frequency of outcome conditional on treatment can be expressed as:

\begin{align}
\begin{aligned}
q_{z|t}&=\frac{\sum_{a\in \mathcal{A}}q_{a,t,z}}{\sum_{a\in \mathcal{A}}q_{t|a}q_a}\\
&=\sum_{a\in \mathcal{A}}q_{z|t,a}q_{a|t}
\end{aligned}
\end{align}
The marginal frequency of outcome conditional on treatment is equal to the sums of the frequencies of outcome conditional on treatment and the confounding variable weighted by the frequency of $a$ conditional on treatment. The source of Simpson's paradox is that different conditional weights can lead to very different conclusions. 

When outcome data are available conditional on a confounder, such as sex in the example above, it is natural to use this information in treatment decisions. Even without direct measurements of potential confounders, we can still include the possibility of their influence in the hypothesis, ensuring that posteriors reflect uncertainty about unobserved factors.

Partial information about confounders is also a common feature of this class of inference problems. For instance, we may know the distribution of sex by treatment, or of outcomes by sex, without the full joint distribution, or we may know that sex and outcome are correlated. While such information is rarely incorporated in standard methods, it is clearly valuable for inference. We show below that constrained entropy-favoring priors provide a transparent way to integrate partial information directly into posterior estimates.

\section{Predictive inference with incomplete information}

In the case of a joint distribution over three variables,  $\mathcal{X}=\{a,t,z\}$, each with two discrete values, the joint distribution is a $2\times2\times2$ tensor array with $8$ occupation numbers, $q_{a,t,z}$ that we refer to as the "more informative" case and present in Table~\ref{table:tensorcontingencytablejoint}. If we have information on the possible confounding variable, we can compute the frequency of outcome conditional on treatment and sex to guide treatment choices.

\setlength{\tabcolsep}{4pt} 
\renewcommand{\arraystretch}{1.2} 
\begin{table}[ht!]
\centering
\caption{Joint frequency distribution $q_{a,t,z}$ as a $2\times2\times2$ tensor.}
\renewcommand{\arraystretch}{1.5}
\begin{tabular}[t]{c||c||c}
\toprule
 & $T$ & $\bar{T}$ \\
\midrule
$A$ & 
\begin{tabular}[t]{c|c}
$Z$ & $q_{A,T,Z}$\\
$\bar{Z}$ & $q_{A,T,\bar{Z}}$
\end{tabular}
&
\begin{tabular}[t]{c|c}
$Z$ & $q_{A,\bar{T},Z}$\\
$\bar{Z}$ & $q_{A,\bar{T},\bar{Z}}$
\end{tabular}\\
\midrule
$\bar{A}$ &
\begin{tabular}[t]{c|c}
$Z$ & $q_{\bar{A},T,Z}$\\
$\bar{Z}$ & $q_{\bar{A},T,\bar{Z}}$
\end{tabular}
&
\begin{tabular}[t]{c|c}
$Z$ & $q_{\bar{A},\bar{T},Z}$\\
$\bar{Z}$ & $q_{\bar{A},\bar{T},\bar{Z}}$
\end{tabular}\\
\bottomrule
\end{tabular}
\label{table:tensorcontingencytablejoint}
\end{table}

It is convenient to write the joint distribution as nested conditionals:
\begin{align}
\label{eq:jointdecomp}
q_{a,t,z}=q_{z|t,a}\,q_{t|a}\,q_{a}.
\end{align}
Each conditional or marginal depends on a single free parameter in $[0,1]$. With three binary variables, there are $7$ independent parameters, matching the degrees of freedom of the normalized joint distribution. Table~\ref{table:tensorcontingencytable} shows the parameterization explicitly.

\begin{table}[ht!]
\centering
\caption{Parameterization of $q_{a,t,z}$ in terms of marginal and conditional frequencies.}
\renewcommand{\arraystretch}{1.5}
\setlength{\tabcolsep}{2pt}
\begin{tabular}[t]{c||c||c}
\toprule
 & $T$ & $\bar{T}$ \\
\midrule
$A$ &
\begin{tabular}[t]{c|c}
$Z$ & $q_{Z|A,T}q_{T|A}q_{A}$\\
$\bar{Z}$ & $(1-q_{Z|A,T})q_{T|A}q_{A}$
\end{tabular}
&
\begin{tabular}[t]{c|c}
$Z$ & $q_{Z|A,\bar{T}}(1-q_{T|A})q_{A}$\\
$\bar{Z}$ & $(1-q_{Z|A,\bar{T}})(1-q_{T|A})q_{A}$
\end{tabular}\\
\midrule
$\bar{A}$ &
\begin{tabular}[t]{c|c}
$Z$ & $q_{Z|\bar{A},T}q_{T|\bar{A}}(1-q_{A})$\\
$\bar{Z}$ & $(1-q_{Z|\bar{A},T})q_{T|\bar{A}}(1-q_{A})$
\end{tabular}
&
\begin{tabular}[t]{c|c}
$Z$ & $q_{Z|\bar{A},\bar{T}}(1-q_{T|\bar{A}})(1-q_{A})$\\
$\bar{Z}$ & $(1-q_{Z|\bar{A},\bar{T}})(1-q_{T|\bar{A}})(1-q_{A})$
\end{tabular}\\
\bottomrule
\end{tabular}
\label{table:tensorcontingencytable}
\end{table}

The posterior probability (Eq.~\ref{eq:multKL}) over the joint frequencies implies a posterior probability over the conditional and marginal frequencies as well. For example, if we are interested in the posterior distribution of recovery for treated males, $\mathcal{P}[q_{Z|A,T}|p_{a,t,z}]$ we need to marginalize over the remaining six dimensions:

\begin{align}
\begin{aligned}
\label{eq:marginalization}
\mathcal{P}&[q_{Z|A,T}|p_{a,t,z}]=\\
&\int_0^1 \cdots \int_0^1 \mathcal{P}[q_{a,t,z}]d q_{A} d q_{T|A} d q_{T|\bar{A}} d q_{Z| A,\bar{T}} d q_{Z| \bar{A},\bar{T}} d q_{Z| \bar{A},T}
\end{aligned}
\end{align}

Doing the same for $\mathcal{P}[q_{Z|A,\bar{T}}|p_{a,t,z}]$ produces a posterior for the frequencies of recovery conditional on treatment and confounding variable that can be used to calculate the posterior treatment effect as the difference in the two densities.  

\subsection{Estimation from Bayesian inference with full posterior distributions}

The reasoning above reveals that there are at least three ways we might estimate the conditional frequency of outcome conditional on sex and treatment, depending on what level of data is available. If the data on clinical trials includes the sex of the subject (Table~\ref{table:SPFull}), we would want to use that information by basing treatment decisions on the posterior probabilities implied by the more informative data. Assuming there are no unmeasured confounding variables we would calculate the posterior over the hypothesis $q_{a,t,z}$ that includes the measured confounding variable conditional on the full information data:
\begin{align}
\label{eq:entfavposteriorhiinfo}
\mathcal{P}[q_{a,t,z}|p_{a,t,z}] \propto e^{H[q_{a,t,z}]}e^{-NH[p_{a,t,z}||q_{a,t,z}]}
\end{align}

From Eq.~\ref{eq:entfavposteriorhiinfo}, the marginal posterior of recovery conditional on treatment and sex, $q_{Z|a,t}$, follows by marginalization (Eq.~\ref{eq:marginalization}).

\begin{figure}[tbhp]
\centering
\includegraphics[width=.47\linewidth]{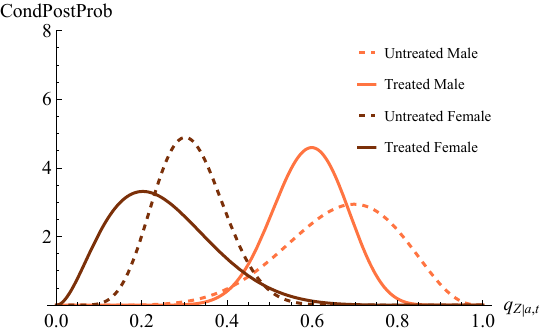} \includegraphics[width=.47\linewidth]{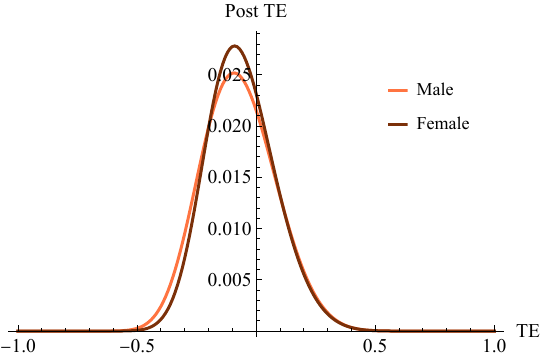}
\caption{Left: Posterior distributions of $q_{Z|A,T},q_{Z|A,\bar{T}},q_{Z|\bar{A},T},q_{Z|\bar{A},\bar{T}}$, with other conditional frequency parameters held at their maximum posterior estimate. The analysis assumes sex is observed and no unmeasured confounders exist. Right: Posterior treatment effect $\tau$ for males and females under the high-information model.}
\label{fig:hiinfopost}
\end{figure}

The posterior treatment effect (PTE) is the distribution of differences in conditional recovery rates,
\[
\tau=q_{Z|T,a}-q_{Z|\bar{T},a}
\]
computed via convolution of the posterior densities:
\[
\mathcal{P}[\tau] = \int_0^1 \mathcal{P}[\tau-q_{Z|T,a}] \, \mathcal{P}[q_{Z|\bar{T},a}] \, dq_{Z|T,a}.
\]

If sex is unobserved (Table~\ref{table:SPLimited}), there are two approaches. The first is to fit a reduced model using only treatment and outcome, effectively assuming no unobserved confounders (equivalent to the assumption of ignorability). This corresponds to a marginalized hypothesis applied to marginalized data:
\begin{align}
\mathcal{P}[q_{a,t,z}|p_{a,t,z},N] \propto e^{H[q_{t,z}]}e^{-NH[p_{t,z}||q_{t,z}]}
\end{align}
We call this the ``low-information model, low-information data'' case.

\begin{figure}[tbhp]
\centering
\includegraphics[width=.47\linewidth]{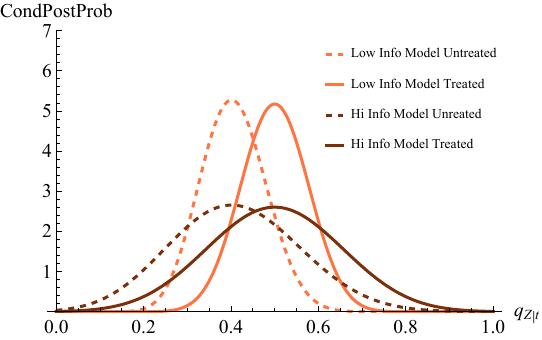} \includegraphics[width=.47\linewidth]{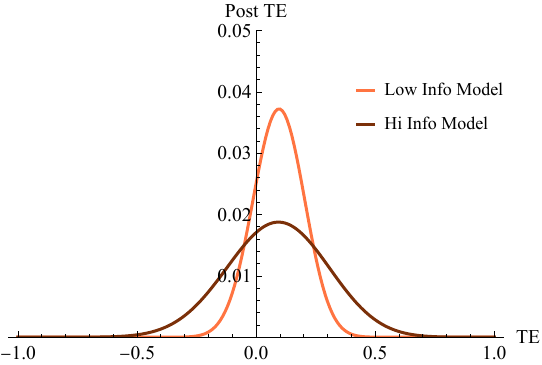}
\caption{Left: Posterior distributions of $q_{Z|A,T},q_{Z|A,\bar{T}},q_{Z|T},q_{Z|\bar{T}}$ for the low-information model (orange curves) and high-information model (brown curves), both fitted to low-information data. Right: Posterior treatment effect $\tau$ for both low-information and high-information models fit to low-information data.}
\label{fig:lowinfopost}
\end{figure}

Alternatively, the hypothesis space can allow for a possible confounder even without direct data. The posterior then becomes:
\begin{align}
\mathcal{P}[q_{a,t,z}|p_{a,t,z}] \propto e^{H[q_{a,t,z}]}e^{-NH[p_{t,z}||q_{t,z}]}
\end{align}
Here the entropy-favoring prior is applied to the full joint distribution, effectively accounting for an unknown confounder. We refer to this as the ``high-information model, low-information data'' case. Figure~\ref{fig:lowinfopost} compares the resulting posteriors for recovery conditional on treatment.

Inspection of Figures~\ref{fig:hiinfopost} and \ref{fig:lowinfopost} reveals key aspects of Simpson's paradox. First, with full information on sex (Table~\ref{table:SPFull}) and assuming no unobserved confounders, the posterior shows treatment is unfavorable for both males and females. 

Second, the low-information model (orange curves, Figure~\ref{fig:lowinfopost}) illustrates Bayesian inference when both the model and data ignore sex (Table~\ref{table:SPLimited}). These posteriors invert the high-information conclusion, showing treatment to be favorable -- a manifestation of Simpson's paradox.

Third, the high-information model (brown curves, Figure~\ref{fig:lowinfopost}) allows for a confounder without direct data. Its posteriors peak at the same conditional frequencies as the low-information case but are much wider, reflecting greater caution. This occurs because entropy-favoring priors assign higher weight to hypotheses that admit possible confounding. 

A clinician using this less restrictive hypothesis would be more cautious about treatment efficacy than one who ignored potential confounding because none was measured. Thus, even without specific data on confounders, Bayesian methods can encode intuitive concerns about data limitations, avoiding overconfident inferences.

\section{Partial Information}
\label{sec:partialinfo}

Another important case arises when partial information beyond outcome and treatment is available. For example, we may know the joint distribution of sex and treatment but not sex and outcome, or vice versa. Such information can still be incorporated by estimating the full model subject to constraints from the partial distribution. To illustrate, we assume knowledge of the joint distribution of sex and outcome but not of sex and treatment. In this case,
\[
\sum_{t\in \mathcal{T}} q_{a,t,z}=\bar{q}_{a,z},
\]
where $\bar{q}_{a,z}$ is obtained by marginalizing Table~\ref{table:SPFull} over $t$. The resulting partial information is shown in Table~\ref{table:paritalinfo}.

\begin{table}[tbhp]
\centering
\caption{Observed joint distribution of sex and outcome, $\bar{q}_{a,z}$, used as partial information when the joint distribution of sex and treatment is unavailable.}
\renewcommand{\arraystretch}{1.5}
\begin{tabular}[t]{c||c||c}
\toprule
 & $Z$ & $\bar{Z}$ \\
\midrule
$A$ & 0.3125 & 0.1875 \\
\midrule
$\bar{A}$ & 0.1375 & 0.3625 \\ 
\bottomrule
\end{tabular}
\vspace{.5cm}
\label{table:paritalinfo}
\end{table}

We incorporate this constraint using the constrained entropy-favoring (CEF) prior, which solves
\begin{align}
\begin{aligned}
\label{eq:maxentIncompInfo}
\max_{q_{a,t,z}} H[q_{a,t,z}]
&= -\sum_{a,t,z} q_{a,t,z}\log q_{a,t,z}, \\
\text{subject to } & \sum_{a,t,z} q_{a,t,z}=1, \\
& \sum_{t} q_{a,t,z} = \bar{q}_{a,z}\quad \forall a,z
\end{aligned}
\end{align}

The Lagrangian for this problem is
\begin{align}
\begin{aligned}
\mathcal{L}[q_{a,t,z},\lambda,\mu_{a,z}]
&= -\sum_{a,t,z} q_{a,t,z}\log q_{a,t,z} \\
&\quad -\lambda\!\left(\sum_{a,t,z}q_{a,t,z}-1\right) \\
&\quad -\sum_{a,z}\mu_{a,z}\!\left(\sum_{t}q_{a,t,z}-\bar{q}_{a,z}\right)
\end{aligned}
\end{align}

The first-order conditions yield the constrained maximum entropy distribution:
\begin{align}
\label{eq:maxentmargpartial}
\hat{q}_{a,z}
=\frac{1}{2}\frac{e^{-\mu_{a,z}}}{\sum_{a,z} e^{-\mu_{a,z}}}
\end{align}
with multipliers $\mu_{a,z}$ determined by the observed $\bar{q}_{a,z}$.  

The posterior incorporating partial information can then be written as the product of two KL divergences:
\begin{align}
\label{eq:entfavposteriorpartinfo}
\mathcal{P}[q_{a,t,z}|p_{a,t,z},\bar{q}_{a,z}]
\;\propto\; e^{-H[q_{a,t,z}\,||\,\hat{q}_{a,z}]} \; e^{-NH[p_{t,z}\,||\,q_{t,z}]}
\end{align}
\begin{figure}[tbhp]
\centering
\includegraphics[width=.47\linewidth]{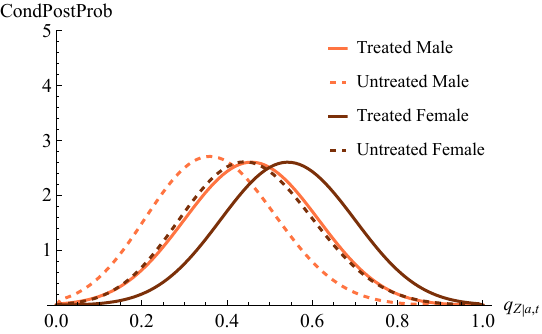}
\includegraphics[width=.47\linewidth]{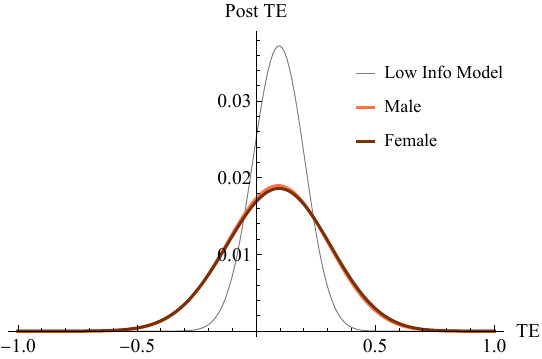}
\caption{Left: Posterior distributions of $q_{Z|A,T},q_{Z|A,\bar{T}},q_{Z|\bar{A},T},q_{Z|\bar{A},\bar{T}}$ under the partial-information model with a CEF prior. Orange curves: males; brown curves: females. All other parameters are fixed at their maximum posterior estimates. Right: Posterior treatment effect $\tau$ under the partial-information model.}
\label{fig:postcomparisonpartial}
\end{figure}

Examining the posterior treatment effect for each sex category (Right Figure~\ref{fig:postcomparisonpartial}) shows that incorporating partial information makes a negative treatment effect considerably more likely -- consistent with the conclusions drawn under full-information analysis.\footnote{While this particular partial information does not noticeably shift the posterior treatment effect, it does affect the posterior distributions $q_{Z|a,t}$. More pronounced impacts of partial information are shown in Figure~\ref{fig:postcomparisonsensitivity} below.}

\section{Sensitivity analysis}

Allowing for unobserved confounders expands the hypothesis space without fixing their direction or magnitude. Sensitivity analysis examines how varying strengths of such confounding affect inferences about treatment effects \citep{Rosenbaum1995}. Within the constrained maximum entropy framework, information can be included as moment constraints, even when their values are unknown \citep{scharfenakerQuantalResponseStatistical2017,FoleyScharfenaker2024a,Scharfenaker2020a,ScharfenakerFoley2024}. One approach is to impose theoretical constraints on the covariance between the unmeasured confounder and both treatment and outcome, capturing how correlations of different magnitudes would alter posterior conclusions.

\begin{align}
\begin{aligned}
\label{eq:maxentIncompInfoSensitivity}
\max_{q_{a,t,z}} H[q_{a,t,z}]
&= -\sum_{a\in \mathcal{A}}\sum_{t\in \mathcal{T}}\sum_{z\in \mathcal{Z}}
q_{a,t,z}\log q_{a,t,z},\\
\text{subject to } & \sum_{a,t,z} q_{a,t,z} = 1,\\
& \text{Cov}[a,t]=\sum_{a,t,z} q_{a,t,z} a t = \alpha,\\
& \text{Cov}[a,z]=\sum_{a,t,z} q_{a,t,z} a z = \delta
\end{aligned}
\end{align}

The Lagrangian is:

\begin{align}
\begin{aligned}
\mathcal{L}[q,\mu,\gamma,\phi]
&=-\sum_{a,t,z} q_{a,t,z}\log q_{a,t,z} \\
&\quad -\mu\!\left(\sum_{a,t,z}q_{a,t,z}-1\right)\\
&\quad -\gamma\!\left(\sum_{a,t,z}q_{a,t,z}at-\alpha\right)\\
&\quad -\phi\!\left(\sum_{a,t,z}q_{a,t,z}az-\delta\right)
\end{aligned}
\end{align}

The first-order conditions imply:

\begin{align}
\label{eq:maxentmarg}
\hat{q}_{a,t,z}
= \frac{e^{-\gamma a z - \phi a t}}
       {\sum_{a,t,z} e^{-\gamma a z - \phi a t}}
\end{align}

Solving for the multipliers under the $\{-1,1\}$ indexing convention (Eq.~\ref{eq:index}) yields:
\begin{align}
\gamma = \tfrac{1}{2}\log\!\left(\frac{1-\alpha}{1+\alpha}\right),
\quad
\phi = \tfrac{1}{2}\log\!\left(\frac{1-\delta}{1+\delta}\right)
\end{align}
so that
\begin{align}
\label{eq:maxentmarg2}
\hat{q}_{a,t,z}
=\tfrac{1}{8}(1+\alpha)\!\left(\frac{1-\alpha}{1+\alpha}\right)^{\tfrac{1}{2}(1-a z)}
(1+\delta)\!\left(\frac{1-\delta}{1+\delta}\right)^{\tfrac{1}{2}(1-a t)}
\end{align}

Because $\alpha,\delta \in [-1,1]$, they are interpretable as the correlation coefficients between the unobserved confounder and treatment or outcome, respectively.  

\begin{figure}[tbhp]
\centering
\includegraphics[scale = .62]{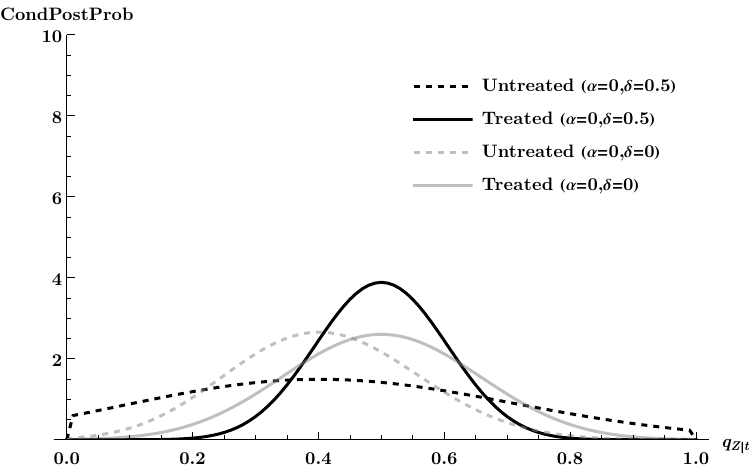} \includegraphics[scale = .62]{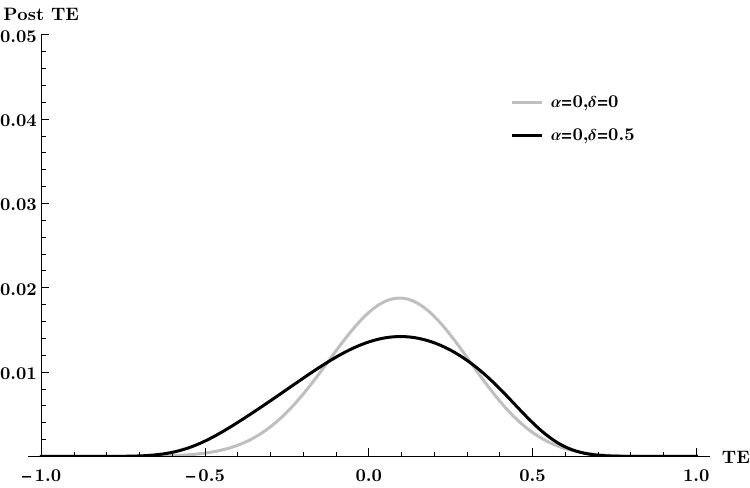}\\
\includegraphics[scale = .62]{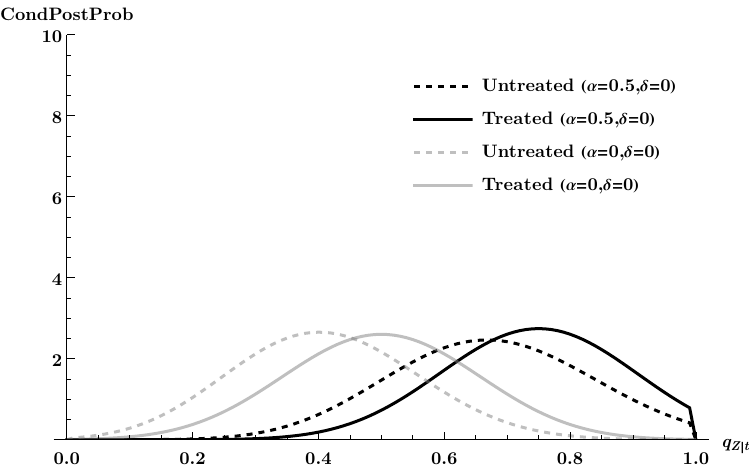} \includegraphics[scale = .62]{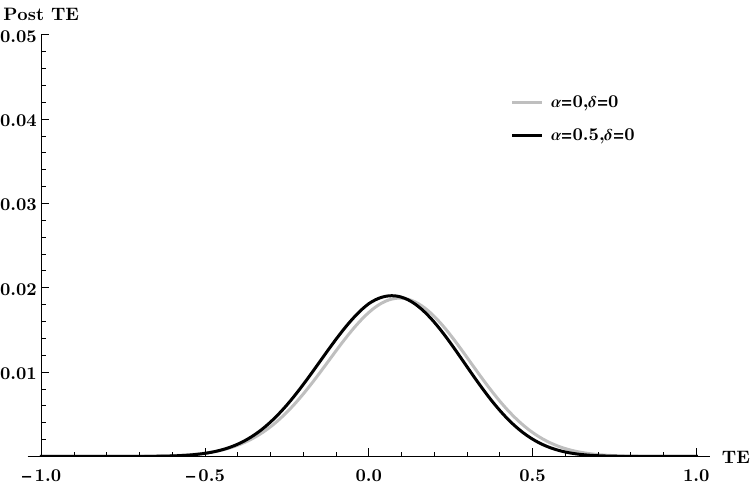}\\
\includegraphics[scale = .62]{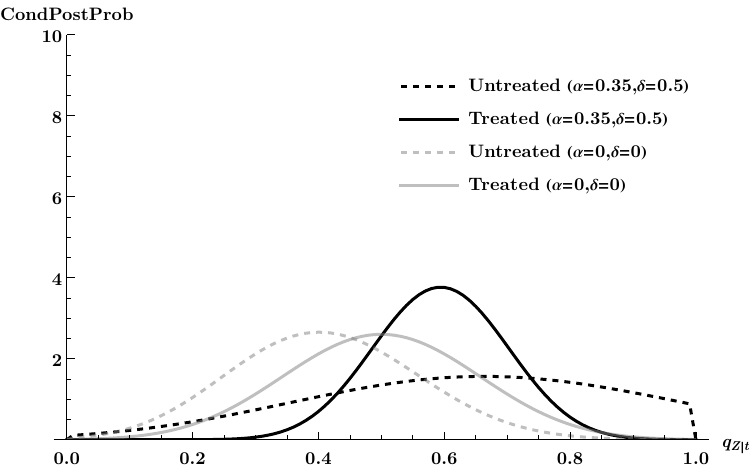} \includegraphics[scale = .62]{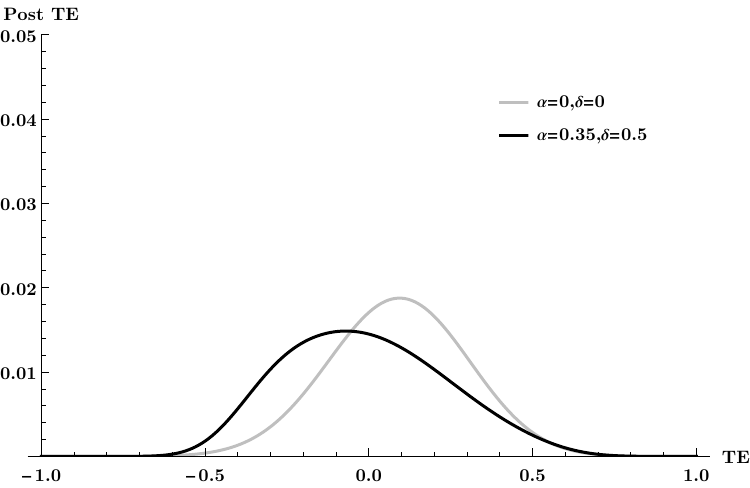}
\caption{Posterior distributions of $q_{Z|T}$ and $q_{Z|\bar{T}}$ under CEF priors with correlation between the unobserved confounder and treatment/outcome. 
Top: $\alpha=0.5$, $\delta=0$. 
Middle: $\alpha=0$, $\delta=0.5$. 
Bottom: $\alpha=0.35$, $\delta=0.5$. 
Gray curves denote the baseline case $\alpha=\delta=0$, identical to the high-information model in Fig.~\ref{fig:lowinfopost}.}
\label{fig:postcomparisonsensitivity}
\end{figure}

Figure~\ref{fig:postcomparisonsensitivity} shows posterior distributions $\mathcal{P}[q_{Z|T}|p_{a,t,z}]$, $\mathcal{P}[q_{Z|\bar{T}}|p_{a,t,z}]$, and the posterior treatment effect under three cases of unmeasured confounding. In the top panel, with $\alpha=0.5$ and $\delta=0$, even strong correlation with treatment has little impact on inference. In the middle panel ($\alpha=0$, $\delta=0.5$), correlation with outcome widens the posterior treatment effect, increasing uncertainty. The bottom panel ($\alpha=0.35$, $\delta=0.5$) reverses the modal treatment effect, resembling the negative effect in the more informative data. These values match empirical expectations, since $\text{Cov}[a,t]=0.35$ and $\text{Cov}[a,z]=0.5$ calculated from Table~\ref{table:SPFull}. With fewer data constraints, uncertainty here is also much greater than in Figure~\ref{fig:hiinfopost}. Crucially, the empirical moment constraint is sufficient to reverse the maximum posterior effect without direct access to the underlying observations, demonstrating the importance of partial information when available.

\section*{Discussion}
The framework developed here generalizes standard econometric approaches to ``causal'' inference by replacing the untestable assumption of ignorability with an explicit probabilistic treatment of latent confounders. In the potential outcomes framework, ``causal'' identification hinges on treatment being independent of potential outcomes conditional on observed covariates \citep{rubin1978}. This assumption restricts the posterior to a subspace that excludes unmeasured dependencies.

In contrast, our Bayesian approach does not limit the hypothesis space to informational structure of the observed data. By incorporating possible unobserved confounders and using constrained entropy-favoring priors over $q_{a,t,z}$, we obtain posterior distributions that reflect the uncertainty inherent in incomplete information. The constrained entropy-favoring prior operationalizes partial exchangeability by assigning greater weight to higher-entropy hypotheses, expressing caution when information is scarce. As a result, posteriors credibly widen in the presence of potential confounders, even when these are unobserved.

This perspective extends the logic of exchangeability. Instead of assuming conditional exchangeability given covariates, we model partial exchangeability directly in the prior. Including a latent confounder relaxes the assumption of full exchangeability without requiring its exact structure, producing more diffuse -- and thus more robust -- posterior distributions of treatment effects.

The approach is also readily extendable to familiar econometric designs. In regression analysis, it allows researchers to model potential confounders directly within the hypothesis space rather than assuming they are absent. When only partial information about the covariance structure of the error term or unobserved regressors is available, constrained entropy-favoring priors provide a principled way to encode this information, ensuring that posterior distributions reflect both observed data and credible uncertainty about latent factors. In difference-in-differences models, for example, the framework relaxes the strict parallel-trends assumption by allowing for latent group-specific heterogeneity, yielding treatment effect posteriors that are typically wider but more credible. When partial information about heterogeneity across groups is available -- such as pre-treatment outcome trends, group-level averages from aggregated data, differences in policy exposure intensity, or structural features like network linkages or geographic proximity -- it can be directly encoded in CEF priors. More broadly, panel regressions, instrumental-variables analyses, and event-study designs can be improved by incorporating partial information about confounding, spillovers, or heterogeneous treatment responses directly into the prior, ensuring that inference remains logically coherent and appropriately cautious.

\section*{Conclusion}

The possibility that unobserved confounding variables may undermine conclusions drawn from data, such as in clinical trials of treatments, has long been a concern in the statistical literature, particularly within the frequentist tradition that has dominated since the 1930s. In recent decades, several authors \citep{Pearl2000,Jensen1996} have proposed ``causal'' models as a response to these problems.

The arguments in this paper show how Bayesian reasoning can incorporate both observed and potentially unobserved confounding variables without relying on the concept of ``causality'', in line with the approach of \cite{lad}. In relation to the literature on ``causal statistical inference'', such models, when framed in Bayesian terms, often take the form of assumed constraints on priors, sometimes represented graphically. As \cite{Lad1999} observes, these constraints can lack clear justifications and occasionally lead to implications that conflict with known features of the scenarios they aim to describe. In many cases, these models seek to introduce some form of the ``ignorability'' assumption -- explicitly or implicitly -- yet this assumption is rarely testable in practice, limiting their ability to resolve concerns about confounding variables.

In Bayesian terms, probability statements represent coherent degrees of belief conditioned on explicit information, and inferences must be consistent with the logic of conditional probability. Predictive inferences thus take the form of posterior conditional distributions that depend on the priors and the information supplied. Our analysis demonstrates how to include the possibility of unobserved confounding variables in the hypothesis space, and how doing so increases posterior uncertainty in treatment effect estimates -- a feature that can help ensure the robustness of conclusions. When additional information is available, such as partial observations or system-level moments, the principle of maximum entropy provides a systematic means of incorporating it into the prior as constraints, yielding posterior inferences that remain logically coherent and appropriately cautious.

In this way, our framework applies Bayesian logic to settings where ignorability and other strong assumptions are not tenable, replacing potentially overconfident conclusions with posterior distributions that better reflect the uncertainty inherent in limited information.

\bibliographystyle{chicago}

\begin{thebibliography}{}

\bibitem[\protect\citeauthoryear{Beni and Liu}{Beni and Liu}{1994}]{Beni1994}
Beni, G. and X.~Liu (1994).
\newblock A least biased fuzzy clustering method.
\newblock {\em IEEE Transactions on Pattern Analysis and Machine Intelligence\/}~{\em 16\/}(9), 954--960.

\bibitem[\protect\citeauthoryear{Chau}{Chau}{2001}]{Chau2001}
Chau, T. (2001).
\newblock Marginal maximum entropy partitioning yields asymptotically consistent probability density functions.
\newblock {\em IEEE Transactions on Pattern Analysis and Machine Intelligence\/}~{\em 23\/}(4), 414--417.

\bibitem[\protect\citeauthoryear{Cover and Thomas}{Cover and Thomas}{2006}]{Cover2006}
Cover, T.~M. and J.~A. Thomas (2006).
\newblock {\em Elements of Information Theory\/} (2 ed.).
\newblock Wiley.

\bibitem[\protect\citeauthoryear{de~Finetti}{de~Finetti}{1974}]{definetti1974}
de~Finetti, B. (1974).
\newblock {\em Theory of Probability: A Critical Introductory Treatment}.
\newblock New York: Wiley.

\bibitem[\protect\citeauthoryear{Flashner-Abramson, Vasudevan, Adejumobi, Sonnenblick, and Kravchenko-Balasha}{Flashner-Abramson et~al.}{2019}]{Flashner2019}
Flashner-Abramson, E., S.~Vasudevan, I.~Adejumobi, A.~Sonnenblick, and N.~Kravchenko-Balasha (2019).
\newblock Decoding cancer heterogeneity: studying patient-specific signaling signatures towards personalized cancer therapy.
\newblock {\em Theranostics\/}~{\em 9\/}(18), 5149--5165.

\bibitem[\protect\citeauthoryear{Foley and Scharfenaker}{Foley and Scharfenaker}{2025}]{FoleyScharfenaker2024a}
Foley, D.~K. and E.~Scharfenaker (2025).
\newblock Bayesian inference and the principle of maximum entropy.
\newblock {\em The American Statistician\/}, 1--7.

\bibitem[\protect\citeauthoryear{Fujino, Ueda, and Saito}{Fujino et~al.}{2008}]{Fujino2008}
Fujino, A., N.~Ueda, and K.~Saito (2008).
\newblock Semisupervised learning for a hybrid generative/discriminative classifier based on the maximum entropy principle.
\newblock {\em IEEE Transactions on Pattern Analysis and Machine Intelligence\/}~{\em 30\/}(3), 424--437.

\bibitem[\protect\citeauthoryear{Golan}{Golan}{2018}]{Golan2018}
Golan, A. (2018).
\newblock {\em Foundations of Info-Metrics: Modeling, Inference and Imperfect Information}.
\newblock New York, NY: Oxford University Press.

\bibitem[\protect\citeauthoryear{Golan and Foley}{Golan and Foley}{2014}]{foleygolan2023}
Golan, A. and D.~K. Foley (2014).
\newblock Understanding the constraints in maximum entropy methods for modeling and inference.
\newblock {\em IEEE Transactions on Pattern Analysis and Machine Intelligence\/}~{\em 45\/}(3), 3994--3998.

\bibitem[\protect\citeauthoryear{Gupta, Gray, and Olshen}{Gupta et~al.}{2006}]{Gupta2006}
Gupta, M., R.~Gray, and R.~Olshen (2006).
\newblock Nonparametric supervised learning by linear interpolation with maximum entropy.
\newblock {\em IEEE Transactions on Pattern Analysis and Machine Intelligence\/}~{\em 28\/}(5), 766--781.

\bibitem[\protect\citeauthoryear{Hardt and Recht}{Hardt and Recht}{2022}]{hardt_recht2022}
Hardt, M. and B.~Recht (2022).
\newblock {\em Patterns, Predictions, and Actions: A story about machine learning}.
\newblock Princeton, NJ: Princeton University Press.

\bibitem[\protect\citeauthoryear{Henter and Kleijn}{Henter and Kleijn}{2016}]{Henter2016}
Henter, G.~E. and W.~B. Kleijn (2016).
\newblock Minimum entropy rate simplification of stochastic processes.
\newblock {\em IEEE Transactions on Pattern Analysis and Machine Intelligence\/}~{\em 38\/}(12), 2487--2500.

\bibitem[\protect\citeauthoryear{Imbens and Rubin}{Imbens and Rubin}{2015}]{rubin2015}
Imbens, G.~W. and D.~B. Rubin (2015).
\newblock {\em Causal Inference for Statistics, Social, and Biomedical Sciences: An Introduction}.
\newblock New York, NY: Cambridge University Press.

\bibitem[\protect\citeauthoryear{Jaynes}{Jaynes}{1957}]{Jaynes_1957b}
Jaynes, E.~T. (1957, may).
\newblock Information theory and statistical mechanics.
\newblock {\em Physical Review\/}~{\em 106\/}(4), 620--630.

\bibitem[\protect\citeauthoryear{Jaynes}{Jaynes}{1968}]{Jaynes1968}
Jaynes, E.~T. (1968).
\newblock Prior probabilities.
\newblock {\em IEEE Transactions On Systems Science and Cybernetics\/}~{\em 4\/}(3), 227--241.

\bibitem[\protect\citeauthoryear{Jaynes}{Jaynes}{1983}]{Jaynes1983}
Jaynes, E.~T. (1983).
\newblock {\em Papers on Probability, Statistics, and Statistical Physics}.
\newblock Reidel.

\bibitem[\protect\citeauthoryear{Jensen}{Jensen}{1996}]{Jensen1996}
Jensen, F.~V. (1996).
\newblock {\em An Introduction to Bayesian Networks}.
\newblock New York, NY: Springer.

\bibitem[\protect\citeauthoryear{Kravchenko-Balasha}{Kravchenko-Balasha}{2020}]{Kravchenko2020}
Kravchenko-Balasha, N. (2020, 10).
\newblock Toward deciphering of cancer imbalances: Using information-theoretic surprisal analysis for understanding of cancer systems.
\newblock In {\em Advances in Info-Metrics: Information and Information Processing across Disciplines}. Oxford University Press.

\bibitem[\protect\citeauthoryear{Lad}{Lad}{1996}]{lad}
Lad, F. (1996).
\newblock {\em Operational subjective statistical methods: a mathematical, philosophical and historical introduction}.
\newblock New York: Wiley.

\bibitem[\protect\citeauthoryear{Lad}{Lad}{1999}]{Lad1999}
Lad, F. (1999).
\newblock Assessing the foundation for bayesian networks: a challenge to the principles and the practice.
\newblock {\em Soft Computing\/}~{\em 3\/}(3), 174--180.

\bibitem[\protect\citeauthoryear{Lezon, Banavar, Cieplak, Maritan, and Fedoroff}{Lezon et~al.}{2006}]{Lezon2006}
Lezon, T.~R., J.~R. Banavar, M.~Cieplak, A.~Maritan, and N.~V. Fedoroff (2006).
\newblock Using the principle of entropy maximization to infer genetic interaction networks from gene expression patterns.
\newblock {\em Proceedings of the National Academy of Sciences\/}~{\em 103\/}(50), 19033--19038.

\bibitem[\protect\citeauthoryear{Lindley and Novick}{Lindley and Novick}{1981}]{Lindley1981}
Lindley, D.~V. and M.~R. Novick (1981).
\newblock The role of exchangeability in inference.
\newblock {\em The Annals of Statistics\/}~{\em 9\/}(1), 45--58.

\bibitem[\protect\citeauthoryear{Murphy and Bassett}{Murphy and Bassett}{2024}]{Murphy2024}
Murphy, K.~A. and D.~S. Bassett (2024, March).
\newblock Information decomposition in complex systems via machine learning.
\newblock {\em Proceedings of the National Academy of Sciences\/}~{\em 121\/}(13), e2312988121.

\bibitem[\protect\citeauthoryear{Novick}{Novick}{1983}]{novick1983}
Novick, M.~R. (1983).
\newblock The centrality of lord's paradox and exchangeability for all statistical inference.
\newblock In H.~Wainer and S.~Messick (Eds.), {\em Principals of Modern Psychological Measurement}, pp.\  41--53. Hillsdale, NJ: Routledge, Earlbaum.

\bibitem[\protect\citeauthoryear{Pearl}{Pearl}{2000}]{Pearl2000}
Pearl, J. (2000).
\newblock {\em Causality: Models, Reasoning, and Inference\/} (1st ed.).
\newblock Cambridge, UK: Cambridge University Press.

\bibitem[\protect\citeauthoryear{Pearl}{Pearl}{2014}]{pearl2014}
Pearl, J. (2014).
\newblock Comment: Understanding {Simpson's Paradox}.
\newblock {\em The American Statistician\/}~{\em 68\/}(1), 8--13.

\bibitem[\protect\citeauthoryear{Qjidaa and Radouane}{Qjidaa and Radouane}{1999}]{Qjidaa1999}
Qjidaa, H. and L.~Radouane (1999).
\newblock { Robust Line Fitting in a Noisy Image by the Method of Moments }.
\newblock {\em IEEE Transactions on Pattern Analysis \& Machine Intelligence\/}~{\em 21\/}(11), 1216--1223.

\bibitem[\protect\citeauthoryear{Remacle, Kravchenko-Balasha, levitzki, and Levine}{Remacle et~al.}{2021}]{Remacle2010}
Remacle, F., N.~Kravchenko-Balasha, A.~L. levitzki, and R.~D. Levine (2021).
\newblock Information-theoretic analysis of phenotype changes in early stages of carcinogenesis.
\newblock {\em PNAS\/}~{\em 107\/}(22), 10324--10329.

\bibitem[\protect\citeauthoryear{Rioux, Scarvelis, Choksi, Hoheisel, and Maréchal}{Rioux et~al.}{2021}]{Rioux2021}
Rioux, G., C.~Scarvelis, R.~Choksi, T.~Hoheisel, and P.~Maréchal (2021).
\newblock Blind deblurring of barcodes via kullback-leibler divergence.
\newblock {\em IEEE Transactions on Pattern Analysis and Machine Intelligence\/}~{\em 43\/}(1), 77--88.

\bibitem[\protect\citeauthoryear{Rosenbaum}{Rosenbaum}{1995}]{Rosenbaum1995}
Rosenbaum, P.~R. (1995).
\newblock {\em Observational Studies}.
\newblock New York: Springer-Verlag.

\bibitem[\protect\citeauthoryear{Rubin}{Rubin}{1978}]{rubin1978}
Rubin, D.~B. (1978).
\newblock Bayesian inference for causal effects: The role of randomization.
\newblock {\em The Annals of Statistics\/}~{\em 6\/}(1), 34--58.

\bibitem[\protect\citeauthoryear{Scharfenaker}{Scharfenaker}{2020}]{Scharfenaker2020a}
Scharfenaker, E. (2020).
\newblock Implications of quantal response statistical equilibrium.
\newblock {\em Journal of Economic Dynamics and Control\/}~{\em 119}, 103990.

\bibitem[\protect\citeauthoryear{Scharfenaker and Foley}{Scharfenaker and Foley}{2017}]{scharfenakerQuantalResponseStatistical2017}
Scharfenaker, E. and D.~Foley (2017, August).
\newblock Quantal {{Response Statistical Equilibrium}} in {{Economic Interactions}}: {{Theory}} and {{Estimation}}.
\newblock {\em Entropy\/}~{\em 19\/}(9), 444.

\bibitem[\protect\citeauthoryear{Scharfenaker and Foley}{Scharfenaker and Foley}{2024}]{ScharfenakerFoley2024}
Scharfenaker, E. and D.~K. Foley (2024).
\newblock Information and entropy in the labor market: Frictional and involuntary unemployment and the neutrality of money.
\newblock {\em Metroeconomica\/}~{\em 76\/}(1), 192--218.

\bibitem[\protect\citeauthoryear{Shin, Remacle, Fan, Hwang, Wei, Ahmad, Levine, and Heath}{Shin et~al.}{2011}]{Young2011}
Shin, Y.~S., F.~Remacle, R.~Fan, K.~Hwang, W.~Wei, H.~Ahmad, R.~Levine, and J.~R. Heath (2011).
\newblock Protein signaling networks from single cell fluctuations and information theory profiling.
\newblock {\em Biophysical Journal\/}~{\em 100\/}(10), 2378--2386.

\bibitem[\protect\citeauthoryear{Skilling and Bryan}{Skilling and Bryan}{1984}]{Skilling1984}
Skilling, J. and R.~K. Bryan (1984, 11).
\newblock Maximum entropy image reconstruction: general algorithm.
\newblock {\em Monthly Notices of the Royal Astronomical Society\/}~{\em 211\/}(1), 111--124.

\bibitem[\protect\citeauthoryear{Smith}{Smith}{1979}]{Smith1979}
Smith, C.~B. (1979).
\newblock A dual method for maximum entropy restoration.
\newblock {\em IEEE Transactions on Pattern Analysis and Machine Intelligence\/}~{\em PAMI-1\/}(4), 411--414.

\bibitem[\protect\citeauthoryear{Zadran, Remacle, and Levine}{Zadran et~al.}{2014}]{Zadran2014}
Zadran, S., F.~Remacle, and R.~Levine (2014, 09).
\newblock Surprisal analysis of glioblastoma multiform (gbm) microrna dynamics unveils tumor specific phenotype.
\newblock {\em PLOS ONE\/}~{\em 9\/}(9), 1--6.

\end{thebibliography}

\end{document}